\documentclass[twocolumn,showpacs,showkeys,preprintnumbers,amsmath,amssymb]{revtex4}
 \usepackage{dcolumn}
 \usepackage{bm}
 \usepackage{graphicx}

\begin{document}

 \title{ Analytical studies of Hawking radiation and
 quasinormal modes in rotating linear dilatonic black hole}

 \author{ Ran Li }

 \thanks{Electronic mail: liran.gm.1983@gmail.com}

 \affiliation{Department of Physics,
 Henan Normal University, Xinxiang 453007, China}

 \begin{abstract}

 The rotating linear dilatonic black hole is an asymptotically non-flat
 solution to Einstein-Maxwell-Dilaton-Axion gravity theory due to
 the existence of non-trivial matter fields. We have analytically
 studied the wave equation of scalar field in this background
 and shown that the radial wave equation can be solved
 in terms of hypergeometric function. By determining the ingoing and the outgoing
 fluxes at the asymptotic infinity, we have found the analytical expressions for
 reflection coefficient and greybody factor for certain scalar modes.
 In the high frequency regime, we obtain the Hawking temperature
 by comparing the blackbody spectrum with the radiation spectrum resulting from
 reflection coefficient. It is shown that the Hawking temperature, which
 depends only on the linear dilatonic background parameter,
 does not agree with the temperature calculated from surface gravity.
 At last, the quasinormal modes of scalar field perturbation are presented,
 which shows that the rotating linear dilationic black hole is unstable
 for certain modes apart from the superradiant modes.

 \end{abstract}
 \pacs{04.70.Dy, 04.62.+v}

 \keywords{dilatonic black hole, Hawking radiation,
 quasinormal modes}

 \maketitle

 \section{Introduction}

 Hawking radiation \cite{hawking} is one of the most significant effects of quantum black holes.
 Although Hawking's original derivation of black hole radiation dealt with the collapsing black
 holes, it is shown by Unruh \cite{unruh} that the eternal black holes also emit particles
 in the form of thermal black body spectrum at the Hawking temperature. The discovery of
 Hawking radiation phenomenon reveals a remarkable connection between thermodynamics, quantum mechanics, and gravity. Due to the significance of this phenomenon, several derivations of Hawking radiation have been proposed in the literatures, including the Damour-Rufini method \cite{damour}, trace anomaly method \cite{trace}, quantum tunneling method \cite{parikh}, gravitational anomaly method \cite{graanomaly1}, etc.

 In the semi-classical approximation, the radiation spectrum of black hole
 can be established by calculating the Bogoliubov coefficients relating
 the vacua near the event horizon and the observer's vacua at the spacial infinity.
 An equivalent procedure is to compute the reflection and the absorption coefficients
 of a wave scattering by the black hole. Usually, the wave equations
 in the background of a general black holes cannot be
 solved exactly, and one must resort to match solutions in an overlap region between the
 near horizon and asymptotic regions.
 The studying of absorption coefficients
 (or greybody factors) by employing the matching technique was set out
 in Refs. \cite{page} and \cite{unruh}. Along this line, the analytical calculation of
 greybody factors and Hawking temperature for various black hole
 backgrounds were studied comprehensively in Refs.\cite{stro,klebanov,cvetic,lee,jung,Fernando,grain,kim,songbai,kk,
 Decanini,Crispino,Gonzalez,ding,cardoso}. Especially,
 the low energy absorption cross section (defined by greybody factor)
 of a minimally coupled massless scalar field in a general spherically
 symmetric black hole exhibits an universal property that the cross section
 is always the area of the horizon \cite{Das}.
 However, in some special cases of the three
 dimensional black holes, such as BTZ black hole \cite{BTZ}, warped $AdS_3$ black hole
 \cite{warpedads} and the self-dual warped $AdS_3$ black hole \cite{selfdualwarped}, the wave equations in such backgrounds can be exactly solved by the hypergeometric function, which allows us to compute
 the reflection and absorption coefficients exactly, leading to the Hawking
 temperatures of these black holes.

 In recent years, it is found \cite{clement1} that the exact semi-classical computation of
 radiation spectrum can also be performed for a class of four dimensional black holes asymptotic to
 linear dilatonic background. This class of black holes, firstly reported by Clement et al.,
 is usually called linear dilatonic black hole, which is a special class of asymptotically
 non-flat black hole solutions to Einstein-Maxwel-Dilaton (EMD) gravity theory.
 In \cite{clement1}, the authors found, in the high frequency regime,
 the temperature for the linear dilaton black holes
 agrees with the surface gravity result. Along the line of \cite{clement1},
 the Hawking radiation spectrum for high dimensional linear dilatonic black holes in the Einstein-Yang-Mills-Dilaton and Einstein-Maxwell-Dilaton theories are carried out in
 \cite{line}, which also shows the agreement of Hawking temperature and the surface gravity
 result.

 More recently, P. I. Slavov and S. S. Yazadjiev \cite{slavov} apply this semi-classical
 approximation method to compute the Hawking radiation of asymptotically non-flat
 dyonic black holes in four dimensional Einstein-Maxwell-dilaton gravity.
 They show that an analytical treatment of the semi-classical radiation
 spectrum of scalar field in non-extremal and extremal black holes
 can be established by solving the wave equations exactly.
 By taking the high frequency limit, the Hawking temperature
 depends only on the linear dilatonic background parameter
 and does not agree with the surface gravity result. This seems to be an
 very interesting result, which motivates us to do the present work.

 In this paper, we will perform an analytical calculation of
 the absorption probability or greybody factor of scalar field
 for a rotating linear dilatonic black hole.
 The rotating linear dilatonic black hole \cite{clement} is an asymptotically non-flat
 solution to Einstein-Maxwell-Dilaton-Axion gravity theory due to
 the existence of non-trivial matter fields. We have analytically
 studied the wave equation of scalar field in the background of
 rotating linear dilatonic black hole and shown that the radial wave equation can be solved
 in terms of hypergeometric function. By determining the ingoing and the outgoing
 fluxes at the asymptotic infinity, we have found the analytical expressions for
 reflection coefficient and greybody factor for certain modes.
 In the high frequency regime, we obtain the Hawking temperature
 by comparing the blackbody spectrum with the radiation spectrum resulting from
 reflection coefficient. It is shown that the Hawking temperature, which
 depends only on the linear dilatonic background parameter,
 does not agree with the temperature calculated from surface gravity.
 We argue that this surprising result is valid for the linear dilatonic
 black holes with the inner and the outer event horizons.
 At last, the quasinormal modes of scalar field perturbation are presented,
 which shows that the rotating linear dilationic black hole is unstable
 for certain modes apart from the superradiant modes.

 This paper is arranged as follows. In Sec.II,
 we give a brief review of the rotating linear dilatonic
 black hole. In Sec.III, we investigate the wave equation
 of massless scalar field and show it is exactly solvable.
 The expressions of Hawking temperature and greybody factor
 are given In Sec.IV. The quasinormal modes and stability
 analysis are presented in Sec.V. The last section is devoted to
 conclusion and discussion.

 \section{Rotating Linear Dilatonic Black Hole}

 In this section, we will briefly review the geometric and thermodynamics properties of
 the rotating linear dilatonic black hole. This class of black hole solution reported by Cl$\acute{\textrm{e}}$ment et al. in Ref.\cite{clement} is derived from
 the Kerr metric by using the solution generating technique. It is a solution
 to Einstein-Maxwell-Dilaton-Axion (EDMA) gravity in four dimensions supported by
 non-trivial gauge field, dilaton field, and axion field.
 At spacial infinity, the rotating linear dilaton black hole is asymptotic
 to linear dilaton background, which is different with Kerr black hole.

 The EMDA gravity theory can be considered arising as a truncated version of the bosonic sector of
 $D=4$, $N=4$ supergravity theory. The action of EMDA gravity theory is given by
 \begin{eqnarray}
 S=\frac{1}{16\pi}\int d^4 x\sqrt{-g}\left[
 -R+2\partial_\mu\phi\partial^\mu\phi
 +\frac{1}{2}e^{4\phi}\partial_\mu\kappa\partial^\mu\kappa
 \right.\nonumber\\
 -\left.e^{-2\phi}F_{\mu\nu}F^{\mu\nu}
 -\kappa F_{\mu\nu}\tilde{F}^{\mu\nu}\right],
 \end{eqnarray}
 where $\phi$ and $\kappa$ are dilaton field and axion (pseudoscalar)
 field respectively, $F$ and $\tilde{F}$ are field strength of
 abelian vector field $A$ and its dual. In \cite{clement}, It is reported that
 the rotating black hole metric without Nut charge is explicitly given by
 \begin{eqnarray}\label{metric}
 ds^2=-\frac{\Delta(r)}{r_0r}dt^2+r_0r \left[\frac{dr^2}{\Delta(r)}
 +d\theta^2+\textrm{sin}^2\theta\big(d\varphi-\frac{a}{r_0r}dt\big)^2 \right]\;,
 \end{eqnarray}
 with the metric function $\Delta(r)=r^2-2Mr+a^2$,
 and the background matter fields are given by
 \begin{eqnarray}
 &&F=\frac{1}{\sqrt{2}}\left[\frac{r^2-a^2\textrm{cos}^2\theta}{r_0r^2}dr\wedge dt
 +a\textrm{sin}2\theta d\theta\wedge\big(d\varphi-\frac{a}{r_0r}dt\big)
 \right]\;, \nonumber\\
 &&e^{-2\phi}=\frac{r_0r}{r^2+a^2\textrm{cos}^2\theta}\;,\nonumber\\
 &&\kappa=-\frac{r_0a\textrm{cos}\theta}{r^2+a^2\textrm{cos}^2\theta}\;.
 \end{eqnarray}

 The two parameters $M$ and $a$ are two integral constants in the metric function,
 which is related to mass and angular momentum of the rotating linear dilatonic black hole.
 The parameter $r_0$ is related to the background electric charge, which represents
 the characteristic of linear dilatonic background.

 The rotating linear dilatonic black hole has two horizons, i.e. the inner and the outer
 event horizons $r_{\pm}$, which are determined by the equation $\Delta(r)=0$.
 The solutions give the locations of event horizons $r_{\pm}=M\pm\sqrt{M^2-a^2}$ providing
 the non-extremal condition $M>a$. The outer and the inner event horizons are
 the coordinate singularities of the metric, which can be eliminated by a proper coordinates transformation.

 Similar to the Kerr black hole case, the rotating linear dilatonic black hole contains
 an ergosphere outside the horizon, which is determined by the equation $r^2-2Mr+a^2\cos^2\theta=0$.
 In asymptotically flat spacetimes with an ergosphere outside the horizon, the scalar field
 will display a superradiant scattering for certain quantum modes, which is a process of transporting the angular momentum of black hole into spacial infinity.
 It is shown in \cite{clement}, because of the non-asymptotically flat property of rotating linear dilatonic black hole, the superradiant modes are reflected from the spacial infinity, i.e.,
 the spacial infinity works as a mirror. The superradiant modes will bounce back and
 forth between the black hole and the mirror. The energy extracted from the black hole
 will gradually grow, which leads to a classical instability of rotating linear dilatonic black hole.

 We now summarize the thermodynamics properties of rotating linear dilatonic black hole.
 It should be noted that $M$ appeared in the solution is no longer the
 physical mass. In order to obtain the first law of black hole
 mechanics, the relevant thermodynamics quantities can be calculated by employing the approach of Brown and York \cite{brown}. The mass $\widetilde{M}$ of rotating linear dilaton black hole
 is relative to the mass parameter $M$ towards the formula
 \begin{eqnarray}
 \widetilde{M}=\frac{M}{2}\;\;,
 \end{eqnarray}
 and the angular momentum $J$ is given by
 \begin{eqnarray}
 J=\frac{ar_0}{2}\;\;.
 \end{eqnarray}
 Here, we firstly present the temperature $T_G$ calculated by the surface gravity.
 The temperature is given by
 \begin{eqnarray}\label{TG}
 T_G=\frac{r_+-r_-}{4\pi r_0 r_+}\;.
 \end{eqnarray}
 It can be shown this temperature satisfies the first law of black hole
 thermodynamics. For this purpose, one can compute the
 Bekenstein-Hawking entropy $S_{BH}$ of black hole
 and the angular velocity $\Omega_H$ of the outer event horizon,
 which are given by
 \begin{eqnarray}
 \Omega_H&=&\frac{a}{r_0 r_+}\;\;,\nonumber\\
 S_{BH}&=&\pi r_0 r_+\;\;.
 \end{eqnarray}

 With the thermodynamical quantities given above,
 one can deduce the differential first law of black hole mechanics
 by straightforward calculation
 \begin{eqnarray}
 d\widetilde{M}=T_GdS_{BH}+\Omega_H dJ\;.
 \end{eqnarray}
 It should be noted that the electric charge $Q=r_0/\sqrt{2}$
 does not appear in the above thermodynamics relation. As mentioned above,
 the parameter $r_0$ is considered as a fixed value. The differentiations
 are performed keeping $Q$ as a fixed value, which is characteristic of the linear dilaton background.

 \section{scalar perturbation: field equation and analytical solution}

 In this section, we will analytically solve the field equation of scalar perturbation
 and study the asymptotic properties of solution near the horizon and spacial infinity.
 For simplicity, we consider the massless scalar field $\Phi$ satisfying the Klein-Gordon
 equation
 \begin{eqnarray}\label{KG}
 \frac{1}{\sqrt{-g}}\partial_\mu\big(\sqrt{-g}g^{\mu\nu}\partial_\nu
 \big)\Phi(t,r,\theta,\varphi)=0\;\;.
 \end{eqnarray}
 After performing the variable separation of scalar field
 $\Phi=e^{-i\omega t}R(r)Y_{lm}(\theta,\varphi)$ and inserting the metric (\ref{metric})
 into Eq.[\ref{KG}], one can obtain the following
 radial equation
 \begin{eqnarray}\label{radialquation}
 &&\partial_r(\Delta\partial_r)R(r)+
 \left(\frac{(\omega r_0r-ma)^2}{\Delta}
 -l(l+1)\right)R(r)=0\;.
 \end{eqnarray}
 Unlike the Kerr case, the angular equation in rotating linear dilatonic black hole
 is just that for the associated Legendre functions.

 In order to solve the radial equation exactly,
 the new variable $z$ should be introduced like this \cite{ranli}
 \begin{eqnarray}
 z=\frac{r-r_+}{r-r_-}\;\;.
 \end{eqnarray}
 Then the radial equation (\ref{radialquation}) can be rewritten in the following
 form after some algebra
 \begin{eqnarray}\label{zeq}
 z(1-z)\frac{d^2
 R}{dz^2}+(1-z)\frac{dR}{dz}+\left(\frac{A}{z}+\frac{B}{1-z}+C\right)R=0\;\;,
 \end{eqnarray}
 where the three parameters $A$, $B$, and $C$ are given by
 \begin{eqnarray}
 A&=&\frac{(\omega r_0 r_+-ma)^2}{(r_+-r_-)^2}\;\;,\nonumber\\
 B&=&\omega^2 r_0^2-l(l+1)\;\;,\\
 C&=&-\frac{(\omega r_0 r_--ma)^2}{(r_+-r_-)^2}\;\;.\nonumber
 \end{eqnarray}

 By redefining the function $R(z)$ as
 \begin{eqnarray}
 R(z)=z^\alpha (1-z)^\beta F(z)\;\;,
 \end{eqnarray}
 with
 \begin{eqnarray}\label{beta}
 \alpha&=&-i\sqrt{A}=-i\frac{\omega r_0 r_+-ma}{r_+-r_-}\;\;,\nonumber\\
 \beta&=&\frac{1}{2}(1-\sqrt{1-4B})
 =\frac{1}{2}-\sqrt{(l+\frac{1}{2})^2-\omega^2r_0^2}\;\;,
 \end{eqnarray}
 the equation (\ref{zeq}) can be transformed to the
 standard hypergeometric function form which is of the form
 \begin{eqnarray}
 z(1-z)F''+(c-(1+a+b)z)F'-abF=0\;\;,
 \end{eqnarray}
 where the parameters $a$, $b$ and $c$ are given by
 \begin{eqnarray}
 a&=&\frac{1}{2}-\sqrt{\left(l+\frac{1}{2}\right)^2-\omega^2 r_0^2}-i\omega r_0\;\;,\nonumber\\
 b&=&\frac{1}{2}-\sqrt{\left(l+\frac{1}{2}\right)^2-\omega^2 r_0^2}-i\frac{\omega r_0(r_++r_-)-2ma}{r_+-r_-}\;\;,\nonumber\\
 c&=&1-\frac{2i(\omega r_0r_+-ma)}{r_+-r_-}\;\;.
 \end{eqnarray}
 Then the general solution of the radial equation (\ref{radialquation}) can be expressed as
 \begin{eqnarray}\label{generalsolution}
 R(z)&=&C_1 z^\alpha (1-z)^\beta F(a,b,c;z)+
 C_2 z^{-\alpha}(1-z)^\beta \nonumber\\
 &&\times F(a-c+1,b-c+1,2-c;z)\;\;.
 \end{eqnarray}

 Now, we consider the asymptotic properties of the solution (\ref{generalsolution}) near the
 horizon and at spacial infinity. Defining the tortoise coordinate near the horizon as
 \begin{eqnarray}
 dr_*=\frac{r_0 r_+}{\Delta}dr\;\;,
 \end{eqnarray}
 which will lead to the relationship $z=e^{2\kappa r_*}$  with
 $\kappa=(r_+-r_-)/2r_0 r_+$ being the surface gravity
 of the outer event horizon. In the limit $z\rightarrow 0$ or $r\rightarrow r_+$,
 by using the property of hypergeometric function $F(a,b,c;0)=1$,
 the solution (\ref{generalsolution}) can be reduced to the form
 \begin{eqnarray}
 R(r\rightarrow r_+)=C_1 e^{-i(\omega-m\Omega_H)r_*}
 +C_2 e^{i(\omega-m\Omega_H)r_*}\;\;.
 \end{eqnarray}
 Therefore, the scalar field $\Phi$ near the horizon behaves as
 \begin{eqnarray}
  \Phi\sim C_1e^{-i\left[\omega t+(\omega-m\Omega_H)r_*\right]}
  +C_2e^{-i\left[\omega t-(\omega-m\Omega_H)r_*\right]}\;.
 \end{eqnarray}
 It is clear that the first term corresponds the ingoing wave while
 the second term is the outgoing wave. In order to match the ingoing
 boundary condition near the horizon, the coefficient $C_2$ should be
 eliminated. Then, we have the general radial solution with the ingoing
 boundary condition at the horizon as
 \begin{eqnarray}\label{finalsolution}
 R(z)=C_1z^\alpha (1-z)^\beta F(a,b,c;z)\;.
 \end{eqnarray}

 To obtain the behavior of the general solution (\ref{finalsolution}) near
 the spatial infinity $z\rightarrow 1$ or $r\rightarrow\infty$,
 the following transformation relationship is useful
 \begin{eqnarray}
 F(a,b,c;z)&=&\frac{\Gamma(c)\Gamma(c-a-b)}{\Gamma(c-a)\Gamma(c-b)}
 F(a,b,a+b-c+1;1-z)\nonumber\\
 &&+(1-z)^{c-a-b}
 \frac{\Gamma(c)\Gamma(a+b-c)}{\Gamma(a)\Gamma(b)}\nonumber\\
 &&\times
 F(c-a,c-b,c-a-b+1;1-z)\;\;.
 \end{eqnarray}
 Performing the transformation leads to
 \begin{eqnarray}
 R(z)&=&C_1 z^{\alpha}\left[(1-z)^\beta
 \frac{\Gamma(c)\Gamma(c-a-b)}{\Gamma(c-a)\Gamma(c-b)}\right.\nonumber\\
 &&\times F(a,b,a+b-c+1;1-z)\nonumber\\
 &&+(1-z)^{1-\beta}\frac{\Gamma(c)\Gamma(a+b-c)}{\Gamma(a)\Gamma(b)}
 \nonumber\\&&\left.\times F(c-a,c-b,c-a-b+1;1-z)\right]\;.
 \end{eqnarray}
 Near the spatial infinity $z\rightarrow 1$ or $r\rightarrow\infty$,
 one can make the approximation for $z$ as
 \begin{eqnarray}
 1-z=\frac{r_+-r_-}{r}\;\;,
 \end{eqnarray}
 which leads to the asymptotic behavior of the radial solution as
 \begin{eqnarray}\label{asymptoticsolution}
 R(r)&=&C_1\left[\left(\frac{r_+-r_-}{r}\right)^\beta
 \frac{\Gamma(c)\Gamma(c-a-b)}{\Gamma(c-a)\Gamma(c-b)}\right.\nonumber\\
 &&\left.
 +\left(\frac{r_+-r_-}{r}\right)^{1-\beta}
 \frac{\Gamma(c)\Gamma(a+b-c)}{\Gamma(a)\Gamma(b)}
 \right]
 \;\;.
 \end{eqnarray}

 Now, we will investigate the radial solution near
 the spatial infinity, where the radial equation (\ref{radialquation}) is reduced to the
 simple form
 \begin{eqnarray}
 \frac{d^2R}{dr^2}+\frac{2}{r}\frac{dR}{dr}+\frac{B}{r^2}=0\;\;.
 \end{eqnarray}
 The solution of the above equation can be easily obtained as
 \begin{eqnarray}\label{asypsolution}
 R(r)=D_1 \left(\frac{1}{r}\right)^\beta+
 D_2 \left(\frac{1}{r}\right)^{1-\beta}\;\;.
 \end{eqnarray}
 Comparing the equation (\ref{asymptoticsolution}) and (\ref{asypsolution}), one can
 obtain the relation between $D_1$, $D_2$ and $C_1$
 \begin{eqnarray}
 D_1&=&C_1(r_+-r_-)^\beta
 \frac{\Gamma(c)\Gamma(c-a-b)}{\Gamma(c-a)\Gamma(c-b)}
 \;,\nonumber\\
 D_2&=&C_1(r_+-r_-)^{1-\beta}
 \frac{\Gamma(c)\Gamma(a+b-c)}{\Gamma(a)\Gamma(b)}
 \;.
 \end{eqnarray}

 Let us summarize briefly the work done in the previous stage. We have considered
 the scalar field perturbation in the background of rotating linear dilatonic
 black hole. After separating the variables, we have obtained an analytical
 solution in terms of hypergeometric function. Generally, in order to study
 the absorption properties of black holes, the boundary condition
 that there are pure ingoing modes near the horizon while there are both ingoing and
 outgoing modes at the spacial infinity should be employed. We have
 studied the asymptotic properties of the analytical solution near the horizon and
 at spacial infinity. It should be noted that, because of the non-asymptotically flat
 property of the rotating linear dilatonic black hole, the distinction between ingoing
 and outgoing fluxes at spacial infinity is a non-trivial task. We will
 discuss this question in the next section.

 \section{Hawking radiation and greybody factor}

 In this section, we will compute the Hawking radiation and greybody factor
 of scalar field for the rotating linear dilatonic black hole.
 Firstly, we should note that the distinction between the ingoing
 and the outgoing fluxes at the asymptotic infinity depends largely on the parameter
 $\beta$. From the expression for $\beta$ in Eq.(\ref{beta}), one can see that
 $\beta$ is real for the low energy modes satisfying the condition $\omega<1/2r_0$,
 while $\beta$ is complex for the high energy modes. In this paper,
 in order to derive the Hawking temperature, we should
 investigate the radiation spectrum in the high energy region.
 For this reason, we will consider the case that the frequency of scalar field
 is larger than $(l+1/2)/r_0$ in this section. The low energy greybody factor
 is presented in the Appendix for completeness.

 For the scalar field modes satisfying the condition $\omega r_0>l+1/2$,
 we should note that the parameter $\beta$ is a complex number. One can analysis the asymptotic solution (\ref{asypsolution}) directly to split the outgoing and the ingoing waves. $\beta$ can be rewritten as
 \begin{eqnarray}
  \beta=\frac{1}{2}-i\tilde{\beta}\;,
 \end{eqnarray}
 where we have introduced a new parameter as $\tilde{\beta}=\sqrt{\omega^2r_0^2-\left(l+1/2\right)^2}$.
 Then the solution (\ref{asypsolution}) can be written in the form of
 \begin{eqnarray}\label{solution}
  R(r)=\frac{D_1}{\sqrt{r}}e^{i\tilde{\beta}\ln{r}}
  +\frac{D_2}{\sqrt{r}}e^{-i\tilde{\beta}\ln{r}}\;.
 \end{eqnarray}
 It is clear that the first term is the outgoing wave while the second term is the ingoing
 wave.

 The conserved flux, up to an irrelevant normalisation coming
 from the angular part, is defined by
 \begin{eqnarray}
 \mathcal{F}=\frac{\sqrt{-g}g^{rr}}{2i}\left(R^*\partial_r R
 -R\partial_r R^*\right)\;.
 \end{eqnarray}
 By substituting the asymptotic equation (\ref{solution}) into the definition of flux, one can obtain
 \begin{eqnarray}
  \mathcal{F}_{asymp}=\frac{\tilde{\beta}}{2}
  \left(|D_1|^2-|D_2|^2\right)\;.
 \end{eqnarray}
 It is clear that $\tilde{\beta}|D_1|^2/2$ represents the outgoing flux
 while $\tilde{\beta}|D_2|^2/2$ represents the ingoing flux.

 The reflection coefficient can be calculated by
 \begin{eqnarray}\label{factor}
  \mathcal{R}&=&\frac{|D_1|^2}{|D_2|^2}\nonumber\\
  &=&\left|\frac{\Gamma\left(\frac{1}{2}-i\left(\tilde{\beta}+\omega r_0\right)\right)\Gamma\left(\frac{1}{2}-i\left(\tilde{\beta}+
  \lambda\right)\right)}
  {\Gamma\left(\frac{1}{2}+i\left(\tilde{\beta}-\omega r_0\right)\right)\Gamma\left(\frac{1}{2}+i\left(\tilde{\beta}-
  \lambda\right)\right)}\right|^2
  \;,
 \end{eqnarray}
 with the parameter $\lambda=\frac{\omega r_0(r_++r_-)-2ma}{r_+-r_-}$.
 By using the Euler¡¯s reflection formula for the Gamma function, one can get
 the final result for the reflection coefficient as
 \begin{eqnarray}\label{factorr}
  \mathcal{R}=\frac{\cosh\pi\left(\tilde{\beta}-\omega r_0\right)
  \cosh\pi\left(\tilde{\beta}-
  \lambda\right)}
  {\cosh\pi\left(\tilde{\beta}+\omega r_0\right)
  \cosh\pi\left(\tilde{\beta}+
  \lambda\right)}
  \;,
 \end{eqnarray}

 For the high frequency modes satisfying the condition $\omega r_0\gg 1$, one can get
 \begin{eqnarray}
  \mathcal{R}\approx \exp(-4\pi\omega r_0)\;.
 \end{eqnarray}
 The reflection coefficient is closely related to the vacuum expectation value of the number operator $N(\omega)$. For bosons, the relation is given by
 \begin{eqnarray}
 \langle N(\omega)\rangle=\frac{\mathcal{R}}{1-\mathcal{R}}.
 \end{eqnarray}
 On the other hand, the vacuum expectation value of the number operator $N(\omega)$
 for the high frequency modes can be approximated by the expression
 \begin{eqnarray}
  \langle N(\omega)\rangle\approx \exp(-\omega/T_H)\;.
 \end{eqnarray}
 One can read off the Hawking temperature of the rotating linear dilatonic black hole as
 \begin{eqnarray}
  T_H=\frac{1}{4\pi r_0}\;.
 \end{eqnarray}

 Comparing this Hawking temperature with the temperature
 computed by the surface gravity shows a notable disagreement.
 This interesting result indicates that the Hawking temperature
 depends only on the linear dilatonic background parameter $r_0$.
 We also find this temperature is coincide with the Hawking
 temperature for the dyonic black hole obtained in \cite{slavov},
 which may be a universal property for the asymptotic linear dilatonic
 black holes with the inner and the outer event horizons.

 Now, we turn to the greybody factor for the rotating linear dilatonic
 black hole. The greybody factor can be calculated by
 \begin{eqnarray}
  \gamma(\omega)=1-\mathcal{R}\;.
 \end{eqnarray}
 The above expression, combined with Eq.(\ref{factor}) or (\ref{factorr}),
 gives the greybody factor for scalar field, valid only in the
 high energy regime. In Fig.1, we depict the greybody factor $\gamma(\omega)$
 for the scalar field modes $l=0,1,2$ and $m=0$. The graphs are
 drawn in such a way that the necessary condition $\omega r_0>l+1/2$ is satisfied.
 It is shown that,
 as the energy parameter $\omega r_0$ increase, the absorption
 probability quickly approaches to unity. This is in agreement
 with the fact that the highly energetic particles always overcome
 the gravitational barrier outside of the horizon of the black
 hole. The high frequency modes are almost swallowed by the black hole.

 \begin{figure}
  \includegraphics[width=0.4\textwidth]{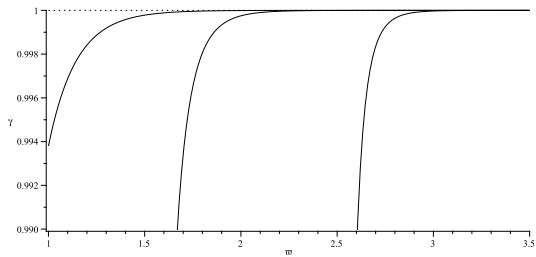}\\
  \caption{Greybody factor $\gamma(\tilde{\omega})$ for the scalar field modes $l=0,1,2$ and $m=0$.
 The graph shift towards right with increasing $l$. The graphs are drawn in such a way that $\tilde{\omega}=\omega r_0>l+1/2$.}
 \end{figure}

 \section{Quasinormal modes and stability analysis}

 In this section, we will compute the quasinormal modes
 by using the analytical solution obtained in Sec.III,
 and analysis the stability of rotating linear dilatonic black hole
 under the scalar field perturbation.

 The quasinormal modes are defined as the modes with the purely ingoing wave
 at the horizon and the purely outgoing wave at the spacial infinity.
 This boundary condition leads to the equation $|D_2|^2=0$.
 So we have the equation
 \begin{eqnarray}
  a=-n\;,\;\;(n=0,1,2,\cdots)
 \end{eqnarray}
 which gives us the explicit expression for the quasinormal modes as
 \begin{eqnarray}
  \omega_n=-\frac{i}{2r_0}\left[n+\frac{1}{2}-
  \frac{(2l+1)^2}{2(2n+1)}\right]\;.
 \end{eqnarray}
 In Fig.2, we have plotted the quasinormal modes for $0\leq n\leq 10$
 and $0\leq l\leq 3$. We can see that the modes satisfying the
 condition $n>l$ are purely damped modes, which indicates the spacetime is stable
 under this kind of perturbation. But, for the modes $n<l$,
 the scalar field perturbation is enlarged along the time, which
 shows that the rotating linear dilatonic black hole
 is unstable. So we have found another kind of instability
 for the rotating linear dilatonic black hole
 apart from the superradiant instability mentioned in Sec.II.

 \begin{figure}
  \includegraphics[width=0.4\textwidth]{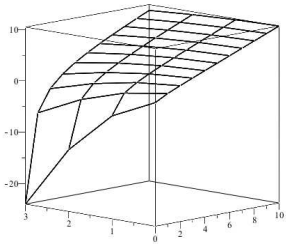}\\
  \caption{Quasinormal modes $i\omega_n r_0$ for $0\leq n\leq 10$
 and $0\leq l\leq 3$.}
 \end{figure}

 \section{conclusion}

 Motivated by the recent study of Hawking radiation of asymptotically non-flat dyonic black holes in four dimensional Einstein-Maxwell-dilaton gravity, we have investigated the corresponding aspect
 for the rotating linear dilatonic black hole in this paper. Benefit from the solvability of
 the radial wave equation in the background rotating linear dilatonic black hole, we can perform an
 analytical calculation of the semi-classical radiation spectrum, greybody factor, and the quasinormal
 modes. By taking the high frequency limit, we have obtained the Hawking temperature
 by comparing the blackbody spectrum with the radiation spectrum resulting from
 reflection coefficient. It is shown that the Hawking temperature, which
 depends only on the linear dilatonic background parameter $r_0$,
 does not agree with the temperature calculated from surface gravity.
 At last, by analyzing the quasinormal modes under the scalar field perturbation,
 we have shown another kind of instability for the rotating linear dilatonic black hole
 apart from the superradiant instability.

  \section*{ACKNOWLEDGEMENT}

 I would like to thank Pu-Jian Mao and Ming-Fan Li for reading the manuscript and useful
 comments. This work was supported by NSFC, China (Grant No. 11147145).

 \section*{Appendix: greybody factor for the modes $\omega r_0<1/2$}

 In the present case, the parameter $\beta$ is real which makes the splitting of the ingoing and the outgoing flux at the asymptotic infinity a non-trivial task. The physical origin is that the rotating linear dilatonic black hole is not asymptotically flat.

 By inserting the asymptotic solution (\ref{asypsolution}) into the definition
 of flux, one can obtain the flux at the asymptotic infinity
 \begin{eqnarray}
  \mathcal{F}_{asymp}=-i\left(\frac{1}{2}-\beta\right)\left(
  D_1D_2^*-D_1^*D_2\right)\;.
 \end{eqnarray}
 We can define the ingoing and the outgoing wave coefficients
 $D_{in}$ and $D_{out}$ by splitting up the coefficients $D_1$ and $D_2$
 as $D_1=D_{in}+D_{out}$ and $D_2=i(D_{in}-D_{out})$.
 Then, the asymptotic flux can be written in the form of
 \begin{eqnarray}
  \mathcal{F}_{asymp}=(1-2\beta)\left(|D_{out}|^2-|D_{in}|^2\right)\;.
 \end{eqnarray}

 The greybody factor can be calculated by
 \begin{eqnarray}
  \gamma(\omega)=1-\frac{|D_{out}|^2}{|D_{in}|^2}
  =\frac{2i(\mathcal{D}-\mathcal{D}^*)}{\mathcal{D}\mathcal{D}^*+i(\mathcal{D}-\mathcal{D}^*)+1}\;,
 \end{eqnarray}
 where $\mathcal{D}$ is given by the following expression
 \begin{eqnarray}
  \mathcal{D}&=&\frac{D_1}{D_2}
  \nonumber\\&=&(r_+-r_-)^{2\beta-1}
 \frac{\Gamma(c-a-b)\Gamma(a)\Gamma(b)}{\Gamma(a+b-c)\Gamma(c-a)\Gamma(c-b)}\;.
 \end{eqnarray}

 One can analyse the properties of greybody factor of scalar field in rotating linear dilatonic  black hole background in the low energy and low angular momentum limits. This aspect
 is out of scope of the present paper.

 \end{document}